\documentclass[preprint,nofootinbib,amsmath,prd,aps,superscriptaddress,tightenlines,12pt]{revtex4}
\usepackage[hypertex]{hyperref}
\usepackage{graphicx}
\usepackage{epstopdf}
\usepackage{bm}
\usepackage{epsfig}
\usepackage{graphics}
\usepackage{xspace}
\preprint{ANL-HEP-PR-10-57}

\def\OMIT#1{}

\newcommand{\nn}{\nonumber}

\newcommand{\bn}{{\bar n}}
\newcommand{\bea}{\begin{eqnarray}}
\newcommand{\eea}{\end{eqnarray}}

\newcommand{\gsim}{\mathrel{\rlap{\lower4pt\hbox{\hskip1pt$\sim$}}\raise1pt\hbox{$>$}}}

\newcommand{\be}{\begin{equation}}
\newcommand{\ee}{\end{equation}}

\begin{document}
\setlength\baselineskip{17pt}

%%%%%%%%%%%%%%%%%%%%%%%%%%%%%%%%%%%%%%%%%%
%Define Title, Author, Address, Preprint#

\title{\bf  Transverse Momentum Distributions in the Non-Perturbative Region}

%\vspace*{1cm}

\author{Sonny Mantry}
\email[]{mantry147@gmail.com}
\affiliation{University of Wisconsin, Madison, WI, 53706}
\author{Frank Petriello}
\email[]{f-petriello@northwestern.edu}
\affiliation{High Energy Physics Division, Argonne National Laboratory, Argonne, IL 60439, USA} 
\affiliation{Department of Physics \& Astronomy, Northwestern University, Evanston, IL 60208, USA}

%\date{\today\\ \vspace{1cm} }

%%%%%%%%%%%%%%%%%%%%%%%%%%%%%%%%%%%%%%%%%%
%Create the title page

\newpage
\begin{abstract}
  \vspace*{0.3cm}
  
We study the low transverse momentum ($p_T$) distribution of the $Z$-boson at hadron colliders
for $p_T \sim \Lambda_{QCD}$ using a factorization and resummation formula derived in the Soft Collinear Effective Theory (SCET).
In the region $p_T \sim \Lambda_{QCD}$, new non-perturbative effects arise that cannot be entirely captured
 by the standard parton distribution functions, and require an additional new non-perturbative transverse momentum function (TMF).  
The TMF is field-theoretically defined in SCET, fully gauge invariant, and captures the 
non-perturbative dynamics that affects the $p_T$-distribution in the region $p_T\sim \Lambda_{QCD}$. The TMF also reduces to the expected 
perturbative result in the region $p_T\gg \Lambda_{QCD}$. We  develop phenomenological models for these TMFs in the non-perturbative region and 
present example fits to the available data.

\end{abstract}

\maketitle

%%%%%%%%%%%%%%%%%%%%%%%%%%%%%%%%%%%%%%%%%%
\newpage
%\tableofcontents
\section{Introduction}

The description of the low transverse momentum ($p_T$)  distribution of electroweak gauge bosons and the Higgs boson has been the subject 
of extensive study~\cite{Dokshitzer:1978yd,Parisi:1979se,Curci:1979bg,Collins:1981uk,Collins:1984kg,Kauffman:1991jt,Yuan:1991we,Ellis:1997ii,
Kulesza:2002rh,Berger:2002ut,Bozzi:2003jy,Kulesza:2003wn,
Bozzi:2005wk, Bozzi:2010xn, Gao:2005iu, Idilbi:2005er}. It plays an important role in the precision measurement of the $W$-boson mass and Higgs boson searches while 
providing an important test of perturbative Quantum Chromodynamics (QCD). In the region of low transverse momentum $p_T \ll M$, where $M$ denotes the mass of 
the electroweak gauge or the Higgs boson, large logarithms of $p_T/M$ spoil the perturbative expansion in the strong coupling and require resummation.

More recently, the low-$p_T$ resummation was studied using a factorization theorem derived~\cite{Mantry:2009qz, Mantry:2010mk} in the Soft-Collinear Effective Theory 
(SCET)~\cite{Bauer:2000yr, Bauer:2001yt, Bauer:2002nz}. The result derived using SCET can be written entirely in momentum space, avoiding issues arising with the 
impact-parameter space present in the standard approach.  All objects in the factorization theorem have well-defined operator expressions.  A detailed 
study of the region $\Lambda_{QCD}\ll p_T \ll M$ was performed for the production of Higgs and 
electroweak gauge bosons. In this region, the factorization theorem is given entirely in terms of perturbatively calculable functions and the standard initial 
state parton distribution functions (PDFs), and takes the schematic form
\bea
\label{schem}
\frac{d^2\sigma}{dp_T^2 \> dY} &\sim & H \otimes {\cal G} \otimes f \otimes f.
\eea
Convolutions between the various objects are denoted by the symbol $\otimes$,  $H$ denotes a hard function whose renormalization group (RG) evolution sums 
logarithms of $p_T/M$, ${\cal G}$ denotes a perturbative function at the $p_T$-scale and describes the emission of soft and collinear partons that recoil 
against the heavy boson, and $f\otimes f$ denotes the product of the initial state PDFs which are evaluated at the $p_T$-scale as 
determined by DGLAP evolution.  
Resummation of the large logarithms was performed at the next-to-leading log (NLL) accuracy in Ref.~\cite{Mantry:2010mk} using 
renormalization group (RG) evolution in the effective theory. The results for the $Z$-boson are in excellent agreement with Tevatron data collected by the 
CDF~\cite{Affolder:1999jh} and D0~\cite{Abbott:1999yd} collaborations.

In this paper, we turn our focus to the region $p_T \sim \Lambda_{QCD} \ll M$ where new non-perturbative effects arise that cannot be captured entirely by the 
standard PDFs. The region $p_T \sim \Lambda_{QCD}$ is sensitive to the transverse momentum distributions of the partons in the initial state hadrons and 
to transverse momentum emissions of order $\Lambda_{QCD}$. In this region, the factorization formula 
takes the schematic form 
\bea
\label{schem2}
\frac{d^2\sigma}{dp_T^2 \> dY} &\sim & H \otimes {\cal K},
\eea
where the function ${\cal K}$ is evaluated at the scale $\mu_T \sim p_T \sim \Lambda_{QCD}$.  The definition of ${\cal K}$ 
is given in Section~\ref{NP}.  It is a new non-perturbative function that cannot be described in terms of the standard PDFs alone. 
In order to facilitate a smooth transition between Eqs.~(\ref{schem2}) and~(\ref{schem}) as one increases $p_T$ from non-perturbative to larger perturbative values, it 
is useful to write ${\cal K}$ in the form
\bea
\label{schem6}
{\cal K} \sim {\cal G} \otimes f \otimes f,
\eea
which defines  the new transverse momentum function (TMF)  ${\cal G}$. For phenomenological purposes, the TMF is modeled in the non-perturbative-$p_T$ region with the constraint that it reduce to the expected perturbative result in Eq.~(\ref{schem}) in the high-$p_T$ region. For this reason we use the same symbol ${\cal G}$ 
to denote the TMF over the entire $p_T$ spectrum.    The 
function ${\cal K}$, or equivalently the TMF ${\cal G}$, is universal and depends only on the hadronic initial state. 

The region of $p_T\sim \Lambda_{QCD}$ has been studied extensively  in the context of semi-inclusive deep-inelastic scattering (SIDIS)~\cite{Collins:1992kk,
Ji:2004wu,Cherednikov:2007tw,Cherednikov:2008ua,Cherednikov:2009wk}, and also within the Collins-Soper-Sterman (CSS) approach to resummation of low-$p_T$ 
logarithms~\cite{Landry:1999an,Qiu:2000hf,Konychev:2005iy}. 
In SIDIS processes, transverse momentum dependent parton distribution functions (TMDPDFs) typically arise in order to describe the order $\Lambda_{QCD}$ dynamics 
in the initial hadrons.
The TMDPDFs are typically not gauge invariant under singular gauge transformations, and arriving at a gauge-invariant definition has been the subject of 
much research~\cite{Collins:1992kk, Ji:2004wu, Ji:2002aa, Cherednikov:2007tw, Cherednikov:2008ua,Cherednikov:2009wk, Idilbi:2010im, Idilbi:2008vm, Bauer:2010cc}. 
In our formalism, it is instead ${\cal K}$ that is the fundamental non-perturbative object in the region $p_T \sim \Lambda_{QCD}$. It 
is fully gauge invariant~\cite{Mantry:2009qz}.  We choose to  write ${\cal K}$ in the form of Eq.~(\ref{schem6}),  and view the TMF (${\cal G}$) and the 
PDFs ($f$) as the fundamental objects of interest. Both of these are manifestly gauge invariant, and have a more intuitive and smoother connection with 
the form of the factorization theorem in the region $p_T\gg \Lambda_{QCD}$. Thus, instead of TMDPDFs it is simpler to  work with the gauge invariant TMFs and 
PDFs which are well-defined over the entire $p_T$-spectrum including both perturbative and non-perturbative values of $p_T$.

The goal of this manuscript is to develop initial models of the TMF that satisfy the following criteria: they reduce smoothly to the perturbative 
result as one increases $p_T$, and they preserve the RG running of ${\cal K}$ in order to cancel the running of the hard function, as required by the scale 
invariance of the cross section.  We do so, and present numerical results for $p \bar{p}$ initial states by fitting to Tevatron data.  A global fit to 
all available data and an analysis of fixed-target data in order to describe the $pp$ initial state and test the universality of the TMF 
is reserved for future work.

The outline of our paper is as follows. In Section~\ref{review}, we review the factorization formulas for the perturbative $p_T$ region 
derived in~\cite{Mantry:2009qz, Mantry:2010mk} and briefly discuss the various pieces and notation.  In Section~\ref{NP} we present the factorization formula for 
the non-perturbative $p_T$-region  and discuss the issues involved in developing a non-perturbative model.  We give phenomenological models for the TMF 
${\cal G}$ in the non-perturbative $p_T$ region and show numerical results in Section~\ref{model}. We conclude in Section~\ref{conclude}.

\section{The Perturbative $\text{p}_\text{T}$ Region}
\label{review}

In this section we briefly review the basic elements of the factorization and resummation formula for the transverse momentum distribution of the 
$Z$-boson in the region $\Lambda_{QCD} \ll p_T \ll M$, as derived in Refs.~\cite{Mantry:2009qz,Mantry:2010mk}. Although we focus on the $Z$-boson, the analysis 
is similar for any color-neutral heavy final state.  The appropriate effective field theory for this observable is SCET$_{\text{II}}$, which has both collinear 
and soft degrees of freedom that can recoil against the $Z$-boson with  transverse momenta of order $p_T$. The collinear and soft degrees of freedom have 
momentum scalings
\bea
p_n \sim M (\eta^2,1,\eta), \qquad p_\bn \sim M (1,\eta^2,\eta), \qquad p_s \sim M (\eta,\eta,\eta), \qquad \eta \sim \frac{p_T}{M_Z},
\eea
where we have used the notation  $p=(n\cdot p,\bn \cdot p, p_\perp)$ to denote the light-cone and transverse momentum components. The light-cone four-vectors 
are $n^\mu =(1,0,0,1)$ and $\bn^\mu=(1,0,0,-1)$. The $p_{n,\bn}$ momenta denote collinear momenta with large components along the $n^\mu$ and $\bn^\mu$ 
directions respectively. The soft momenta are denoted by $p_s$. The transverse momentum distribution in the region $\Lambda_{QCD} \ll p_T \ll M$ is 
dominated by these collinear and soft  modes radiated from the initial state partons. In SCET$_{\text{II}}$, these emissions build up into collinear and soft 
Wilson lines that dress the $Z$-production current. The final factorization and resummation formula for the differential cross-section of the $Z$-boson as a 
function of its transverse momentum and rapidity (Y) is given by
\bea
\label{intro-DY}
\frac{d^2\sigma}{dp_T^2\> dY}&=& \frac{\pi^2 }{N_c^2}   \int_0^1 dx_1 \int_0^1 dx_2\int_{x_1}^1 \frac{dx_1'}{x_1'} \int_{x_2}^1 \frac{dx_2'}{x_2'}   \nn \\
&\times&   H_Z^{q}(x_1x_2Q^2,\mu_Q;\mu_T) \>{\cal G}^{qrs}(x_1,x_2,x_1',x_2',p_T,Y,\mu_T) f_r(x_1',\mu_T)  f_s(x_2',\mu_T).\nn \\  
\eea
The above formula involves a convolution of three types of objects: the hard function $H_Z^q$, the TMF ${\cal G}^{qrs}$, 
and the initial state PDFs $f_{r,s}$. The indices $r,s$ run over the initial partons and the superscript $q$ denotes the fact that the Z-boson production vertex involves a quark current. The hard function describes the physics of modes with virtuality $p^2\sim M^2$ that are integrated out at the 
scale $\mu_Q \sim M$. The hard function is then evolved down to the scale $\mu_T \sim p_T$ via its renormalization group equations, summing large logarithms 
of order $M/p_T$ in the process. The TMF function ${\cal G}^{qrs}$ lives at the $\mu_T \sim p_T$ scale and describes the physics of the soft and 
collinear emissions in a way that is consistent with the constraints imposed on the $p_T$ and $Y$ of the $Z$-boson. The initial state PDFs $f_{r,s}$ are evaluated at the 
$\mu_T$ scale after DGLAP evolution from the non-perturbative scale, summing logarithms of order $\Lambda_{QCD}/p_T$ in the process.

The TMF function ${\cal G}^{qrs}$ has the form
\bea
\label{intro-DY-2}
 &&{\cal G}^{qrs}(x_1,x_2,x_1',x_2',p_T,Y,\mu_T)=  \int \frac{d^2b_\perp}{(2\pi)^2} J_0\big [b_\perp p_T\big ]\>\int dt_n^+ dt_\bn^- \> 
{\cal I}_{n;q r}(\frac{x_1}{x_1'}, t_n^+,b_\perp,\mu_T)\>\nn \\
&\times&{\cal I}_{\bn;\bar{q} s}(\frac{x_2}{x_2'}, t_\bn^-,b_\perp,\mu_T) {\cal S}^{-1}_{qq}(x_1 Q-e^{Y}\sqrt{\text{p}_T^2+M^2}-\frac{t_\bn^-}{x_2 Q}, 
x_2 Q-e^{-Y}\sqrt{\text{p}_T^2+M^2}- \frac{t_n^+}{x_1 Q},b_\perp,\mu_T),\nn \\
\eea
where the functions ${\cal I}_{n;q r}, {\cal I}_{\bn;\bar{q} s}$ correspond to collinear emissions in the $n$ and $\bn$ directions respectively and 
${\cal S}$ correspond to soft emissions. The inverse soft function (iSF) ${\cal S}^{-1}$ arises due to zero-bin subtractions~\cite{Mantry:2009qz,Mantry:2010mk, Manohar:2006nz,Lee:2006nr, Idilbi:2007ff,Idilbi:2007yi}
necessary to avoid the problem of double-counting the soft region. The collinear function ${\cal I}_{n;q r}$ is defined through the matching of a nucleon 
matrix element called  the impact-parameter beam function (iBF) $\tilde{B}_n^q$ onto the standard PDFs as
\begin{equation}
\tilde{B}_{n}^q(x, t,b_\perp,\mu_T) \equiv \int _x^1\> \frac{dz}{z}\> {\cal I}_{n; qr} (\frac{x}{z}, t,b_{\perp}, \mu_T) \>f_{r}(z,\mu_T),
\label{iBFmatch}
\end{equation}
with an analogous equation for the $\bn$-sector. For  precise field-theoretic definitions of the iBFs, the iSF, and the hard-function $H_Z^q$ we refer the 
reader to Refs.~\cite{Mantry:2009qz,Mantry:2010mk}. Analogous nucleon beam functions~\cite{Fleming:2006cd,Stewart:2009yx, Stewart:2010pd, Stewart:2010qs} are known to appear in other collider processes.  We note that this SCET formalism accomplishes the resummation of large logarithms differently than the 
traditional QCD approach.  Logarithms of the matching-scale ratio $\text{ln} (\mu_Q / \mu_T)$ that appear in the partonic cross section are resummed 
via the RG evolution of $H_Q^Z$.  Upon identifying $\mu_Q \sim M$ and $\mu_T \sim p_T$, these become the standard small-$p_T$ logarithms.  Kinematic logarithms 
which directly have $\text{ln} (M / p_T)$ appear after integration over the momentum fractions $x^{'}_{1,2}$ in Eq.~(\ref{intro-DY}).  It was shown in 
Ref.~\cite{Mantry:2010mk} that this formalism reproduces the correct logarithms upon expansion of the resummed result to the fixed order ${\cal O}(\alpha_s^2)$ given our current knowledge of ${\cal G}$ in 
perturbation theory.

We comment briefly on the recent work of Ref.~\cite{Becher:2010tm} which also uses SCET to address $p_T$-resummation. We disagree with several aspects of their 
results. Their analysis is based on the claim that the emission of soft  radiation with transverse momentum of order $p_T$ does not affect the spectrum of the 
$Z$-boson. This is in 
contrast to our effective field theory (EFT) where both collinear and soft radiation, with transverse momentum of order $p_T$, play a dynamical role in 
determining the transverse 
momentum spectrum. 
It is well-known~\cite{Bauer:2000yr, Bauer:2001yt, Bauer:2002nz} that the emission of multiple collinear and soft partons from the initial-state collinear 
partons build into  eikonal Wilson lines and is a leading order effect in  SCET$_{\text{II}}$. 
Since Ref.~\cite{Becher:2010tm} argues against the presence of effects from soft radiation, their factorization formula does not have the analogue 
of the iSF.  In our formalism the combined RG running of the two iBFs and the iSF cancels the running of the hard function as required by RG invariance. 
The presence of the iSF, which itself has a non-zero anomalous dimension, plays a crucial role in achieving this RG invariance as was shown in 
Ref.~\cite{Mantry:2009qz}. Since Ref.~\cite{Becher:2010tm} does not have the iSF they do not naturally achieve the required RG invariance. 
Instead,  RG invariance is implemented by introducing a `hidden' $Q^2$ dependence in their two nucleon beam functions, which are individually ill-defined. 
This hidden $Q^2$ dependence is argued to arise from a collinear anomaly due to the absence of soft modes.  It is further stated 
that soft modes have a vanishing contribution if the collinear anomaly is properly 
regularized.  Since the presence of the collinear-anomaly already assumes the absence of soft modes, we do not find this argument compelling. 
The hidden $Q^2$ dependence  reappears in their matching of the nucleon beam functions onto the PDFs at the $p_T$-scale, despite the fact that $p_T^2 \ll Q^2$. 
All of these problems are avoided if one starts with the correct degrees of freedom and includes the effects of both 
collinear and soft radiation.  SCET$_{\text{II}}$ is known to be the appropriate EFT for this purpose.

\section{The Non-Perturbative  $\text{p}_\text{T}$ Region}
\label{NP}

In the previous section we reviewed the factorization formula for the region $\Lambda_{QCD}\ll p_T \ll M$. We now consider the $p_T$ distribution in the 
region where $p_T\sim \Lambda_{QCD}$.  The factorization theorem is given by
\bea
\label{intro-DY-NP}
\frac{d^2\sigma}{dp_T^2\> dY}&=& \frac{\pi^2 }{N_c^2}   \int_0^1 dx_1 \int_0^1 dx_2     H_Z^{q}(x_1x_2Q^2,\mu_Q;\mu_T) \>{\cal K}^{q}(x_1,x_2,p_T,Y,\mu_T),\nn \\  
\eea
where ${\cal K}^{q}$ is defined as
\bea
\label{Kq}
{\cal K}^{q}(x_1,x_2,p_T,Y,\mu_T) &\equiv& \int dt_n^+ \int dt_\bn^- \int \frac{d^2b_\perp}{(2\pi)^2} J_0(b_\perp p_T) \tilde{B}_n^q(x_1,t_n^+,b_\perp,\mu_T) 
\tilde{B}_\bn^{\bar{q}}(x_2,t_\bn^-,b_\perp,\mu_T) \nn \\
&\times& {\cal S}^{-1}_{qq}(x_1 Q - e^Y \sqrt{p_T^2+ M^2} - \frac{t_\bn^-}{x_2Q}, x_2Q - e^{-Y} \sqrt{p_T^2 +M^2}- \frac{t_n^+}{x_1Q},b_\perp, \mu_T). \nn \\
\eea
In this case, the iBFs ($\tilde{B}^q_{n,\bn}$) and the iSF (${\cal S}_{qq}^{-1}$) are evaluated at the scale $\mu_T \sim p_T \sim \Lambda_{QCD}$ with the hard function 
$H^q_Z$  evolved via its RG equations 
down to this same scale. Since $\mu_T\sim \Lambda_{QCD}$, the iBFs and the iSF are non-perturbative. This expression for ${\cal K}^q$ was already derived in 
Ref.~\cite{Mantry:2009qz, Mantry:2010mk}. In that work we focused on the region $p_T\gg \Lambda_{QCD}$ so that the iBFs and iSF were perturbative and the iBFs were 
further matched onto PDFs. In this case, since  $\mu_T \sim p_T \sim \Lambda_{QCD}$, the iBFs and iSF are non-perturbative.  A perturbative matching onto PDFs is 
no longer valid and the final form of the factorization theorem is given by Eqs.~(\ref{intro-DY-NP}) and~(\ref{Kq}). For phenomenological purposes, the non-perturbative function ${\cal K}^q$ must be modeled. When $p_T \gg \Lambda_{QCD}$, the scale $\mu_T\sim p_T$ is perturbative.  ${\cal K}^{q}$ then becomes 
a perturbative object and the iBFs can be matched onto PDFs as in Eq.~(\ref{iBFmatch}), leading to Eq.~(\ref{intro-DY}). 

We model the function ${\cal K}^q$ by imposing two requirements. First, the model for ${\cal K}^q$ must preserve the correct RG evolution properties so 
that it cancels the running of the hard function $H^q_Z$, as required by the scale invariance of the cross section. Second, as one increases $p_T$ from the 
non-perturbative region to higher perturbative values, Eq.~(\ref{intro-DY-NP}) must reduce to Eq.~(\ref{intro-DY}).  In order to smoothly transition between the 
non-perturbative and perturbative values of $p_T$, we write the the iBFs in Eq.~(\ref{Kq}) as in Eq.~(\ref{iBFmatch}), even in the non-perturbative region 
where $\mu_T \sim p_T\sim \Lambda_{QCD}$. 
In this region, Eq.~(\ref{iBFmatch}) is no longer a perturbative matching equation but instead 
defines a new non-perturbative function ${\cal I}_{n;qr}$. As one increases $\mu_T\sim p_T$ to perturbative values, the function ${\cal I}_{n;qr}$ corresponds to the perturbatively calculable coefficient in the matching of the iBF onto the PDF. Similar statements apply to the $\bn$-sector iBF. With these conventions, one can write the 
function ${\cal K}^q$ as
\bea
\label{Kqrs}
{\cal K}^{q}(x_1,x_2,p_T,Y,\mu_T) &\equiv& \sum _{r,s}\int _{x_1}^1 \frac{dx_1'}{x_1'} \int _{x_2}^1 \frac{dx_2'}{x_2'} \> {\cal G}^{qrs}(x_1,x_2,x_1',x_2', 
p_T,Y,\mu_T)f_r(x_1',\mu_T)f_s(x_2',\mu_T),\nn \\
\eea
for \textit{all} values of $p_T$.  For perturbative $p_T$-values, the quantity ${\cal G}^{qrs}$ is perturbative and identical to that given in Eq.~(\ref{intro-DY-2}). The 
expression in Eq.~(\ref{intro-DY-NP}) then properly reduces to the factorization theorem of Eq.~(\ref{intro-DY}), valid in the region 
$\Lambda_{QCD}\ll p_T \ll M_Z$. For $\mu_T \sim p_T\sim \Lambda_{QCD}$, ${\cal G}^{qrs}$  is non-perturbative, as are all the quantities on the RHS of 
Eq.~(\ref{intro-DY-2}).  In the non-perturbative-$p_T$ region, ${\cal G}^{qrs}$ can be interpreted as the non-perturbative TMF controlling the dynamics of 
transverse momentum dynamics of order $\Lambda_{QCD}$.  The modeling of the function ${\cal K}^q$ in this region is
reduced to the modeling of the TMF ${\cal G}^{qrs}$. 

We view Eq.~(\ref{Kqrs}) with ${\cal K}^q$ rewritten in terms of the standard initial state PDFs and a new TMF function ${\cal G}^{qrs}$ as more convenient 
than the form in Eq.~(\ref{Kq}). Both ways of writing ${\cal K}^q$ are equally valid. In Eq.~(\ref{Kq}), the iBFs might be associated 
with TMDPDFs in the language used for the study of SIDIS processes.  These iBFs are invariant under covariant gauge transformations but are not in general 
invariant under singular  gauge transformations. However, the full product of the two iBFs and iSF that define ${\cal K}^q$ is completely gauge invariant. 
For a more detailed discussion of this point we refer the reader to Ref.~\cite{Mantry:2009qz}. The form of ${\cal K}^q$ in Eq.~(\ref{Kqrs}) makes  gauge invariance 
manifest.  Since both ${\cal K}^q$ and the PDFs are gauge invariant, 
the TMF ${\cal G}^{qrs}$ is also gauge independent. In Eq.~(\ref{iBFmatch}), the gauge dependence of the iBF 
$\tilde{B}_n^q$ under singular gauge transformations is isolated into the function ${\cal I}_{n;qr}$. 
This situation also applies to the $\bn$-sector iBF.  The gauge dependence of the iBFs then cancels in the product that defines ${\cal G}^{qrs}$. In this way, 
the non-perturbative dynamics in the region $p_T\sim \Lambda_{QCD}$ is  described in terms of gauge invariant initial state PDFs and the TMF.

%%%%%%%%%%%%%%%%%%%
\section{TMF Models and Numerical Results}
\label{model}

In this section we develop phenomenological models for the TMF function ${\cal G}^{qrs}$ in the non-perturbative region. We require that the model for 
${\cal G}^{qrs}$ reduces to the perturbatively calculable result as one increases $p_T$. We write ${\cal G}^{qrs}$ in the form
\bea
\label{mod1}
{\cal G}^{qrs}(x_1,x_2,x_1',x_2',p_T,Y,\mu_T) &=& \int _0^\infty dp_T' \>{\cal G}^{qrs}_{\text{part.}}(x_1,x_2,x_1',x_2', p_T\sqrt{1+(p_T'/p_T)^2}, Y,\mu_T)\nn \\
&\times&\> G_{mod}(p_T', a,b, \Lambda), \nn \\
\eea
which is a convolution of the partonic result for the TMF function ${\cal G}^{qrs}_{\text{part.}}$ with a model function 
$G_{mod}$~\cite{Fleming:2007qr,Hoang:2007vb,Fleming:2007xt}.  This form is reminiscent of that used in the CSS approach, where the integrand of the 
Fourier transform is decomposed according to
\begin{equation}
W(b) = W(b_{*}) W^{NP}(b), \;\; b_{*} = \frac{b}{\sqrt{1+(b/b_{max})^2}}.
\end{equation}
$W(b)$ is the perturbative resummed contribution and $W^{NP}$ denotes the non-perturbative contribution. $b_{max}$ is a free parameter 
typically taken to be of order $1 \, {\rm GeV}^{-1}$.

We parametrize our non-perturbative contribution as
\bea
\label{gmod}
G_{mod}(p_T',a,b, \Lambda) &=& \frac{N}{\Lambda^2} \Bigg (\frac{p_T'^{\>2}}{\Lambda^2}\Bigg )^{a-1} \text{exp} \Big [- \frac{(p_T'-b)^2}{2\Lambda^2}\Big],
\eea
and fix $N$ by the normalization condition
\bea
\label{norm}
\int _0^\infty dp_T' \> G_{mod}(p_T', a,b,  \Lambda) &=& 1.
\eea
In principle, the model function $G_{mod}$ can have flavor indices $r,s$. For the sake of simplicity we will work with a flavor-independent model function $G_{mod}$.
Different choices of the parameters $a,\kappa, \Lambda$ correspond to different model choices for the non-perturbative TMF ${\cal G}^{qrs}$. The model function
parameters are chosen such that $G_{mod}$ will peak at $p_T'\sim \Lambda_{QCD}$ with an exponential fall off for larger values of $p_T'$. As a result, 
${\cal G}^{qrs}$ in Eq.~(\ref{mod1}) receives sizeable contributions only from the region $p_T'\sim \Lambda_{QCD}$. Thus, in the region $p_T\gg \Lambda_{QCD}$ 
one can Taylor expand ${\cal G}^{qrs}_{part.}$ around the limit $p_T\gg p_T'\sim \Lambda_{QCD}$. When combined with Eq.~(\ref{norm}) this gives
\bea
\label{GOPE}
{\cal G}^{qrs}(x_1,x_2,x_1',x_2',p_T,Y,\mu_T)\Big |_{p_T\gg \Lambda_{QCD}} &=& {\cal G}^{qrs}_{\text{part.}}(x_1,x_2,x_1',x_2', p_T, Y,\mu_T) 
+ {\cal O}(\frac{\Lambda_{QCD}}{p_T}). \nn \\
\eea
In the region of perturbative $p_T$, the function ${\cal G}^{qrs}$ properly reduces to its perturbative limit with all model dependence suppressed by 
powers of $\Lambda_{QCD}/p_T$.  In this way, the model dependence is restricted to the non-perturbative region, as expected. 
The perturbative region of the $p_T$ spectrum remains calculable in a model-independent way to leading order in $\Lambda_{QCD}/p_T$. One could consider more 
sophisticated model functions that contain $x$-dependence and that incorporate additional effects, but we restrict ourselves in this initial analysis 
to the form of Eq.~(\ref{gmod}).

\begin{figure}
\includegraphics[height=6in,width=4in,angle=90]{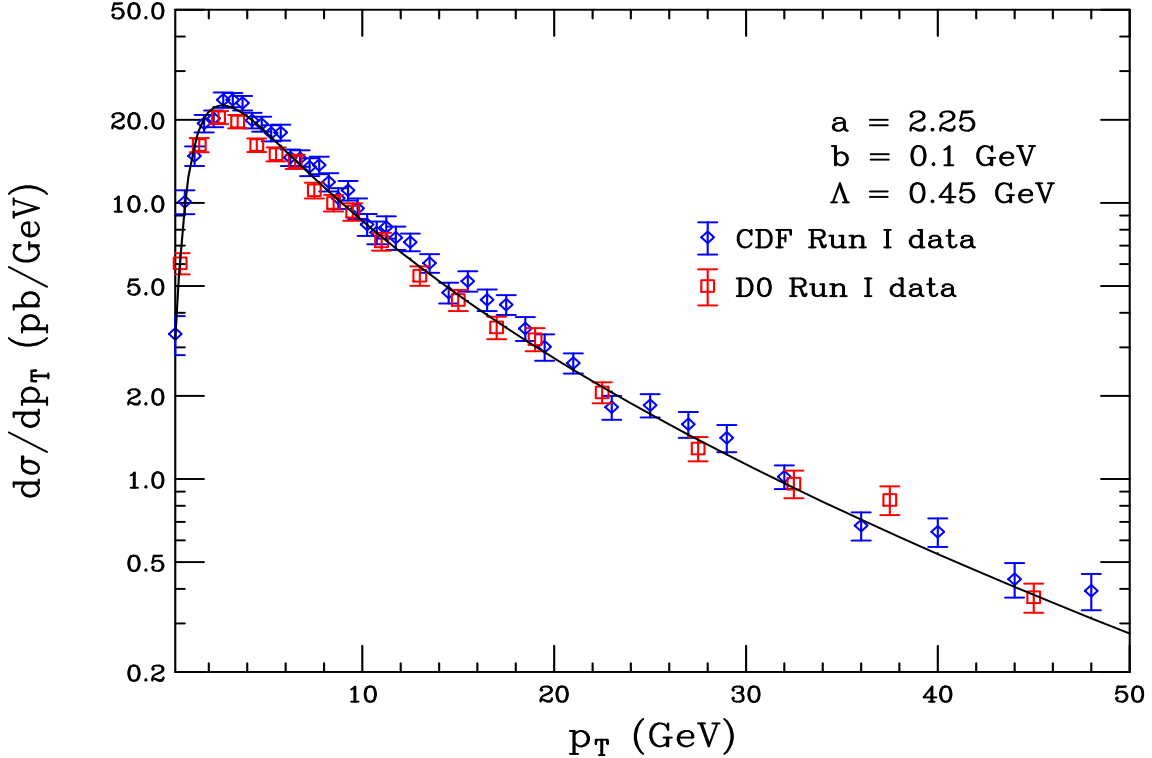}
\caption{The result for the $p_T$-spectrum of the Z-boson for the best fit parameter choices $a=2.25, b=0.1 \,\text{GeV}, \Lambda=0.45\,\text{GeV}$. We have also 
set $\mu_Q^2 = -M_Z^2$ and $\mu_T^2 = p_T^2 +p_{Tmin}^2$ where $p_{Tmin}= 1$ GeV. The data points were collected by the CDF and D0 
collaborations~\cite{Affolder:1999jh,Abbott:1999yd}.}
\label{fig:ZpT2}
\end{figure}

The implementation of the model also requires care regarding the choice of the scale  $\mu_T$. In the perturbative $p_T$ region, the scale 
$\mu_T \sim p_T$ is the appropriate choice. However, one cannot use $\mu_T \sim p_T$ when $p_T$ is of order $ \Lambda_{QCD}$ or smaller. The RG 
equations for the evolution of the hard function $H^q_Z(x_1 x_2Q^2,\mu_Q;\mu_T)$ become non-perturbative in this region, and
${\cal G}^{qrs}_{part.}$ in Eq.~(\ref{mod1}) becomes incalculable. A sensible choice for $\mu_T$ that can be applied in both the perturbative and 
non-perturbative $p_T$ regions is
\bea
\label{muTmod}
\mu_T^2 = \xi^2 \> p_T^2 + p_{Tmin}^2,
\eea
where $p_{Tmin}\gsim 1$ GeV is a low, but still perturbative, scale and can be viewed as another parameter of the model.  It is analogous to the parameter 
$b_{max}$ that appears in the CSS approach to transverse momentum resummation.  $\xi$ is a scale variation parameter 
we take to be ${\cal O}(1)$. The above choice of scale for $\mu_T$ has several useful properties. As $p_T\to 0$, the scale $\mu_T\to p_{Tmin}$ 
so that ${\cal G}^{qrs}_{part}$ in Eq.~(\ref{mod1}) is still evaluated at a perturbative scale. Similarly, the running of the hard function 
$H_Z^{q}(x_1x_2Q^2,\mu_Q;\mu_T)$ will freeze at the perturbative scale $p_{Tmin}$ as $p_T\to 0$. For larger values of $p_T\gg p_{Tmin}\gsim 1$ GeV, 
$\mu_T \to \xi \> p_T$ so that the appropriate choice of $\mu_T \sim p_T$ in the perturbative region is recovered.

\label{num}
\begin{figure}
\includegraphics[height=6in, width=4in, angle=90]{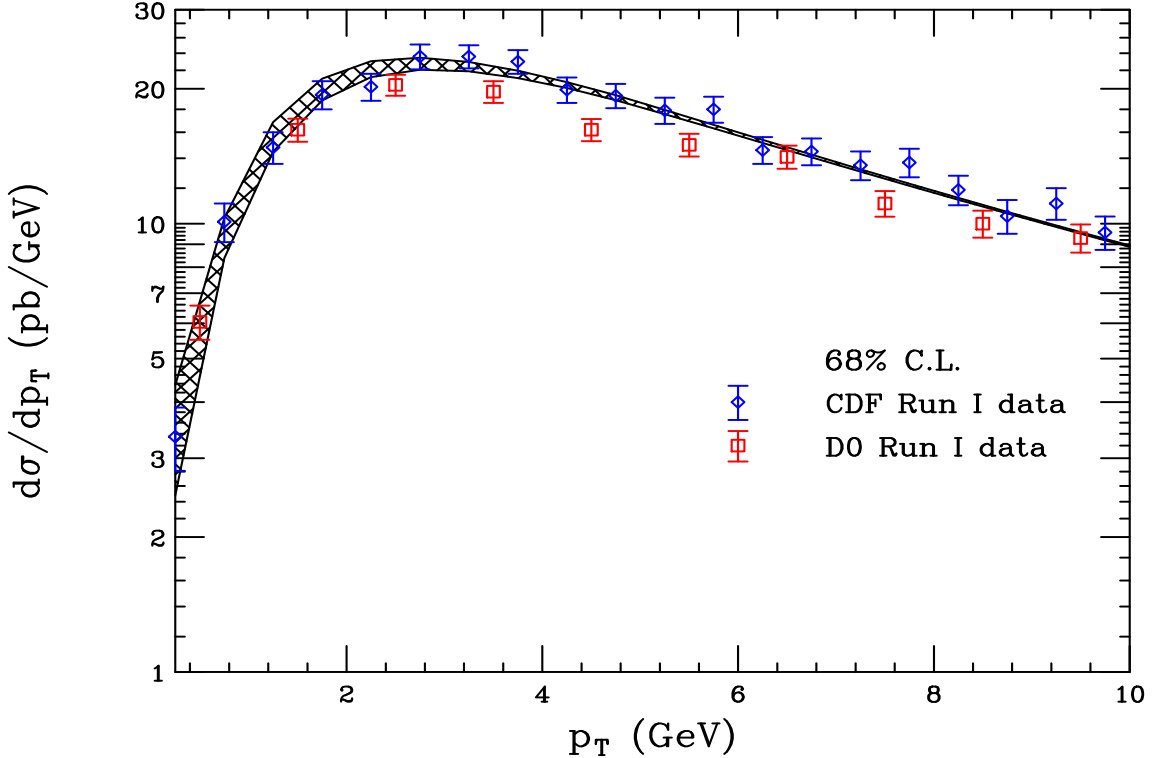}
\caption{The result of varying the model parameters $a,b$, and $\Lambda$ within their 68\% confidence level allowed region. We have chosen $\mu_Q^2=-M_Z^2$, 
$\mu_T^2 = p_T^2 +p_{Tmin}^2$ with $p_{Tmin}=1$ GeV. We see that the variation of the model parameters only affects the very low $p_T$ region and has a 
negligible effect in the region $p_T \gg \Lambda_{QCD}$. The data points are from the CDF and D0 collaborations~\cite{Affolder:1999jh,Abbott:1999yd}.}
\label{fig:ZpT3}
\end{figure}

We now present an example fit of the TMF function  ${\cal G}^{qrs}$ to Tevatron data for the $Z$-boson $p_T$ spectrum.  We choose 
$\mu_Q^2 = -M_Z^2$~\cite{Ahrens:2008qu,Ahrens:2008nc}, $\mu_T$ as in Eq.~(\ref{muTmod}) with $\xi=1$, and for simplicity set $p_{Tmin}=1$ GeV.  We note that 
this ensures that the scale $\mu_T$ at which the PDFs are evaluated always remains at or above the initial scale $Q_0 = 1$ GeV used in the 
MSTW fit~\cite{Martin:2009iq}, 
a criterion pointed out in previous work in the CSS approach~\cite{Konychev:2005iy}.  We then 
perform a chi-squared fit of the parameters $a,b$, and $\Lambda$ in Eq.(\ref{gmod}) against CDF data~\cite{Affolder:1999jh}; for simplicity we do not 
include the D0 data in this example fit.  The 
best fit values obtained are $a=2.25,b=0.1\,\text{GeV},\Lambda=0.45\,\text{GeV}$ with a goodness-of-fit measure $\chi^2/d.o.f. \sim 0.7$. 
The result for these best fit values are shown in 
Fig.~\ref{fig:ZpT2} along with the CDF and D0 data points. Fig.~\ref{fig:ZpT2} shows that the TMF model is flexible enough to give a good 
description of data in the region $p_T < 1$ GeV where non-perturbative transverse momentum dynamics becomes important.  At the same 
time, a good description of the data is also achieved for larger perturbative values of $p_T$ where the result is given in terms of 
a perturbatively calculable TMF function.  The model dependence introduced by $G_{mod}$ turns off in the region 
$p_T\gg \Lambda_{QCD}$, as expected. This is further illustrated in Fig.~\ref{fig:ZpT3} where we show the results for the 68\% confidence level region in the parameters $a,b$, and $\Lambda$. 
We see in Fig.~\ref{fig:ZpT3} that while the different parameter choices affect the $p_T$-distribution in the non-perturbative region, there is almost no 
effect in the region $p_T \gg \Lambda_{QCD}$. This is a reflection of Eq.~(\ref{GOPE}) which shows that for $p_T \gg \Lambda_{QCD}$ the model dependence is 
power suppressed and the TMF function reduces to the expected partonic result.

Before concluding we comment briefly on the universality of $G_{mod}$.  We have neglected the possible flavor dependence of this function in our fit, indicating 
that we expect the non-perturbative dynamics of the valence up and down quarks that dominate $Z$-boson production at the Tevatron to be the same.  However, it remains 
to be seen whether the valence-sea scattering which occurs in $pp$ collisions can be described by the same $G_{mod}$.  For this reason we refrain from 
making predictions for LHC production until this universality is tested by a detailed fit to the available data.  We note that $W^{NP}$ in the CSS approach 
has been found to satisfy the universality assumption~\cite{Konychev:2005iy}.

\section{Conclusions}
\label{conclude}
In this manuscript we have performed an initial analysis of the $Z$-boson transverse momentum distribution in the region $p_T \sim \Lambda_{QCD}$ 
using a factorization and resummation theorem derived in SCET. Combined with our previous work~\cite{Mantry:2009qz, Mantry:2010mk} which focused on the 
region $\Lambda_{QCD} \ll p_T \ll M$, a description of the entire $p_T$-spectrum is now achieved in the framework of SCET. 
This formalism is free of the Landau poles that arise in the traditional approach to low-$p_T$ resummation in impact-parameter space, and are therefore independent 
of ambiguities and numerical difficulties which arise when transforming back to momentum space. 
In the region where $p_T\sim \Lambda_{QCD}$, the transverse momentum spectrum is affected by new non-perturbative effects that cannot be described by the 
standard PDFs alone. 
A new transverse momentum function (TMF), fully gauge invariant and defined in SCET, arises in addition to the standard PDFs. The TMF captures the 
non-perturbative dynamics associated with the initial state transverse momentum distributions and with final-state emissions having 
transverse momenta of order $\Lambda_{QCD}$. We have devised phenomenological models for the TMFs in the region $p_T\sim \Lambda_{QCD}$. These models are 
such that the TMF reduces to the expected perturbative result when $p_T \gg \Lambda_{QCD}$. This allows for a smooth transition 
between the  non-perturbative and perturbative values of $p_T$. The TMF models also have the correct renormalization group evolution properties built in. We 
have given example fits of the TMF model to Tevatron data. The results of the fit for the TMF function give a good description of 
the CDF and D0 data over the entire $p_T$ spectrum. 

The work presented here is simply the first step in understanding the non-perturbative transverse momentum region within SCET.  A more global analysis of the 
available data is left to future work, as is the modeling of the TMF for $pp$ initial states.  In principle, the TMF is different for $pp$ and $p\bar{p}$ initial 
states.  The universality of this function remains to be studied.  These questions must be addressed to present predictions for the $p_T$ distribution at 
the LHC.  We look forward to these future investigations.

\acknowledgments{This work is supported in part by the U.S. Department of Energy, Division of High Energy Physics, under contract DE-AC02-06CH11357 and the 
grants  DE-FG02-95ER40896 and DE-FG02-08ER4153, and by Northwestern University.
}
\bibliographystyle{h-physrev3.bst}
\bibliography{tmf}

\end{document}